\documentclass[preprint,prd]{revtex4}

\usepackage{amsbsy}
\usepackage{amssymb}
\usepackage{amsmath}
\usepackage{graphicx}

\def\X{{\mathrm{x}}}
\def\x{{\mathrm{x}}}
\def\y{{\mathrm{y}}}

\def\Y{{\mathrm{y}}}

\def\s{{\mathrm{s}}}
\def\N{{\mathrm{N}}}

\def\cW{{\mathcal{W}}}

\def\cS{{\mathcal{S}}}

\def\p{{\rm p}}
\def\e{{\rm e}}

\def\s{\mathrm{s}}

\newcommand{\A}{{\cal A}}
\newcommand{\B}{{\cal B}}

\newcommand{\R}{{\cal R}}

\def\x{{\mathrm{x}}}
\def\y{{\mathrm{y}}}

\def\x{{\mathrm{x}}}
\def\y{{\mathrm{y}}}

\def\p{{\rm p}}
\def\e{{\rm e}}

\def\e{{\rm e}}
\def\s{{\rm s}}

\def\N{{\mathrm N}}

\def\be{\begin{equation}}
\def\ee{\end{equation}}
\def\beq{\begin{equation}}
\def\eeq{\end{equation}}
\def\bea{\begin{eqnarray}}
\def\eea{\end{eqnarray}}

\def\bear{\begin{eqnarray}}
\def\eear{\end{eqnarray}}

\begin{document}

\title{Resistive relativistic magnetohydrodynamics from a  charged multi-fluids perspective}

\author{N. Andersson}

\affiliation{
School of Mathematics, University of Southampton,
Southampton SO17 1BJ, United Kingdom}

\begin{abstract}
We consider general relativistic magnetohydrodynamics from a charged multifluids point-of-view, taking a variational approach as our starting point. We develop the case of two charged components in detail, accounting for a phenomenological resistivity,  providing specific examples for pair plasmas and proton-electron systems. We discuss both cold, low velocity, plasmas and hot systems  where we account for a dynamical entropy component. The results for the cold case (which accord with recent work in the literature) provide a complete model for resistive relativistic magnetohydrodynamics, clarifying the assumptions that lead to various models that have been used in astrophysical applications. The analysis of the hot case is (as far as we are aware) novel,  accounting for the relaxation times that are required to ensure causality 
and  demonstrating the explicit coupling between fluxes of heat and charge.
\end{abstract}

\maketitle

\section{Introduction}

Magnetic fields are ubiquitous in the Universe, affecting physics across a vast range of scales. The relevance of electromagnetism for our everyday experience is obvious. Electromagnetic effects are also central to many processes in astrophysics and cosmology. The strongest known magnetic fields (above $10^{14}$~G) are found in a subclass of neutron stars aptly referred to as magnetars \cite{dynamo,magnetar}, systems that also form the largest (and hottest!) known superconductors \cite{casa1,casa2}. Magnetic fields are equally relevant on the vastly larger scale of entire galaxies, and are likely to have played a role in the early Universe as well \cite{ellis,ellis2,barrow}. Understanding the origin and evolution of electromagnetic fields in their many different guises remains a fundamental question for modern science. 
The literature on the subject is (understandably) vast~\footnote{In view of this, the reference list will be rather incomplete and focused on either recent work or contributions of particular relevance for the discussion.}, yet some problems remain relatively unexplored. This paper concerns one such problem.

Our aim is to develop a model for resistivity in general relativistic magnetohydrodynamics. By necessity this forces us to consider a charged multi-fluid system (we obviously need charged components in relative motion in order to have a charge current!). This part of the problem is quite straightforward; we  make progress by marrying the standard variational model for electromagnetism \cite{efsth} to the charged fluid version of Brandon Carter's convective variational description of relativistic fluids \cite{carter,livrev}. Adding a phenomenological resistivity to the mix is  not difficult, either.  Combining these ingredients we follow the text-book strategy \cite{schnack,bellan} and derive the simplified equations of magnetohydrodynamics. As long as we limit the analysis to low velocities (cold plasmas) the results follow readily. We demonstrate this for the particular problem of a two-component system, composed either of protons and electrons or a pair-plasma with positrons and electrons, and compare our results to the recent literature \cite{koide08,koide09,tsagas}.

The complexity of the problem increases significantly if we turn our attention to high velocities and hot plasmas. One reason for this is obvious: In order to describe a hot system we need to allow for the presence of heat flow. However, the problem of heat in relativistic systems is known to be difficult, as a naive implementation  inevitably leads to causality violation and unwanted instabilities \cite{is1,is2,hislin1,hislin2}. We avoid falling into this trap by building on a recent model that treats the entropy as an additional ``fluid'', which couples to the substantial matter components through entrainment \cite{carterdiss,cesar1,cesar2}. This effect represents the inertia of heat, and leads to the thermal relaxation that is required to ensure causality and stability. The presence of this coupling makes the analysis less straightforward, and the final results are (obviously) less transparent than in the low-velocity case. However, they are also more ``interesting''. The more complicated setting allows for a number of additional features, most notably a coupling between the heat flow and the charge current. From a fundamental point of view one would expect such a thermo-electric coupling \cite{extro}, but this effect has nevertheless not been previously discussed in a relativistic context. 

\section{Charged relativistic multi-fluid systems}

This section sets the stage for the discussion by bringing together and adapting established results from the literature. The key building blocks are obvious: We need 
a framework for discussing electromagnetism in general relativity, and in order to understand the nature of the associated current we also need a 
multi-fluid formulation for charged components. The first part can be found in many standard text-books (see for example \cite{efsth}). The multi-fluid part is less mainstream fare, but
the required formalism (mainly designed by Carter and colleagues, see \cite{carter,livrev} for reviews) has been developed to the required level. The marriage of the two systems has not been discussed extensively in the literature but, as we will see, it is comfortable.

\subsection{Variational multi-fluid dynamics}

Multi-fluid dynamics arise whenever a system has several components, each in the ``fluid regime'', which retain their identity. 
The archetypal such system, known to be well described by a two-fluid model, is superfluid $^4$He \cite{helium}. In principle, one can imagine systems where the 
mean-free path due to inter-species scattering is much larger than that for intra-species scattering \cite{formalism}. On intermediate scales one can then 
meaningfully discuss different fluid components. This set-up may seem somewhat artificial, but there clearly are systems in nature 
where this separation of scales occurs. One reason why superfluid systems tend to require a multi-fluid approach is that the relevant scale 
deciding the ``size of the fluid elements'' is not the mean-free path (since particle scattering is suppressed in a superfluid) but the 
coherence length of the relevant condensate. This scale is usually much smaller than the mean-free path in the corresponding system at  temperatures above the superfluid transition,  so the system ends up acting as a fluid on 
much smaller length scales than usual. In an astrophysical context, the modelling of mature neutron star cores  must account for superfluidity (and superconductivity!).
Indeed, most applications of the general relativistic multi-fluid formalism have been in that problem area, see for example \cite{lin}. 

The model we consider builds on the convective variational principle developed by Carter \cite{carter}. This method  deals, in a natural way, with
the fact that
a variational derivation of the equations of fluid dynamics must be constrained. The development takes as starting point a Lagrangian for the 
matter, $\Lambda$, which is built from all relevant fluxes $n_\X^a$ in the system~\footnote{Throughout the discussion, different fluid components are labelled 
by a constituent index $\X,\Y,\ldots$. The Einstein summation convention for repeated indices does not apply to these. In contrast, the summation 
convention obviously does apply to spacetime indices, which will be represented by italics $a,b,c,\ldots$.}. In the variational approach, the 
conservation of the individual fluxes;
\begin{equation} 
    \nabla_{a} n^{a}_\X = 0 \ , \label{consv2} 
\end{equation} 
is ensured by means of a pull-back construction based on the notion of a three-dimensional matter space. This exercise identifies the 
spacetime  displacements $\xi_\X^a$ that guarantee \eqref{consv2}, and with respect to which the variation of the Lagrangian is carried out. The detailed 
procedure is discussed in \cite{livrev}. For later convenience we simply note that the final result is
\be
\delta n_\X^a = n_\X^b \nabla_b \xi_\X^a - \xi_\X^b \nabla_b n_\X^a - n_\X^a \left( \nabla_b \xi_\X^b + {1\over 2} g^{bc} \delta g_{bc} \right)  \ ,
\label{dna}\ee
where $g_{ab}$ is the spacetime metric and $\delta g_{ab}$ is the induced variation.

A key strength of the variational approach is that it correctly identifies the momentum $\mu^\X_a$ that is conjugate to each flux.
This is crucial in a multi-fluid system since the momenta should encode the so-called entrainment effect \cite{livrev}. As an illustration of how this effect arises, consider a general 
isotropic Lagrangian. Taking the view that the fluxes are the fundamental variables in the problem, we can build this Lagrangian from 
the different scalars that we can construct. This means that we should consider both
\be
n_\x^2 = - n_\x^a n^\x_a \ , 
\ee
which defines the number density of the x component, 
and
\be
n_{\x\y}^2 = - n_\x^a n^\y_a \ , \qquad \y \neq \x \ .
\ee
An unconstrained variation of $\Lambda$ with respect to the independent 
vectors $n^a_\X$ and the metric $g_{ab}$ then leads to
\begin{equation} 
    \delta \Lambda = \sum_{\X} \mu^\X_a \delta n^a_\X + 
                     \frac{1}{2} g^{cb} \left(\sum_{\X} n^a_\X \mu_c^\X\right) \delta 
                     g_{ab} \ , 
\end{equation} 
where the  momenta are given by
\begin{equation} 
    \mu^\X_{a} = g_{ab} \left(\B^{\X } n^b_\X + \sum_{\y\neq\x}\A^{\X \Y} 
                   n^b_\Y\right) \ , \label{mudef2} 
\end{equation} 
with coefficients
\be
\B^\x = -  2 \frac{\partial \Lambda}{\partial   n^2_{\X}} \ , 
\ee
and
\be 
    \A^{\X \Y} = \A^{\Y \X} = - \frac{\partial \Lambda}{\partial 
                 n^2_{\X \Y}} \quad , \quad \X \neq \Y \ . \label{coef12} 
\ee 
The momenta are  dynamically, and 
thermodynamically, conjugate to their respective number density fluxes, and their magnitudes are the chemical potentials (as we will see later).
The $\A^{\X \Y}$ coefficients represent the fact that each fluid momentum  $\mu^\X_{a}$ 
may, in general, be given by a linear combination of the individual 
currents $n^{a}_\X$. \ That is, the current and momentum for a particular 
fluid do not have to be parallel. \ This is  the entrainment effect.
In the limit where all the currents are parallel, e.g. when the fluids are 
comoving, $- \Lambda$ corresponds to the local thermodynamic 
energy density, but in the general case this is not so. 

In terms of the constrained Lagrangian displacements, $\xi_\X^a$, a variation of 
$\Lambda$  yields~\footnote{Here, and in the following, we omit terms that can be written as a total divergence and which lead to surface terms after integration. In principle, these terms can be removed either by appealing to boundary conditions or via the introduction of additional constraints. This is standard procedure discussed in many textbooks.}
\be
    \delta \left(\sqrt{- g} \Lambda\right) = \frac{1}{2} \sqrt{- g} 
    \left(\Psi \delta^a{}_b + \sum_{\X } n^{a}_\X 
    \mu_b^\X\right) g^{bc} \delta g_{ac} -  
    \sqrt{- g} \sum_{\X} f^\X_{a} \xi^{a}_\X 
    \ , 
\label{vars}\ee
where we have defined
\be
f^\X_a =  2 n_\x^b  \nabla_{[b} \mu^\x_{a]}  \ , 
\ee
(and the square brackets indicate anti-symmetrization, as usual).
It follows immediately that the equations of motion for the individual fluids are expressed as
an integrability condition on the vorticity (associated with the momentum not the flux!);
\be
f^\X_a = 0 \ .
\label{fas}\ee 
In \eqref{vars} we have also introduced the generalized pressure 
$\Psi$, defined by 
\begin{equation}
     \Psi = \Lambda - \sum_{\X } n^{a}_\X \mu^\X_{a} \ . 
\end{equation} 

Finally, we want to account for the coupling between the matter flow and the dynamics of spacetime~\footnote{A significant part of the literature on astrophysical magnetic fields has focussed on fixed spacetimes, e.g. in the context of black-hole or neutron-star magnetospheres or jet dynamics. In the present discussion we do not assume that the spacetime dynamics is frozen, even though we do not explicitly discuss the Einstein field equations and the metric degrees of freedom. The aim is to keep the discussion at a sufficiently general level that it can be applied to problems where the live gravitational field plays a relevant role.}. The coupling to gravity follows readily from the fact that the 
stress-energy tensor is obtained as the variation of the matter Lagrangian with respect to the
spacetime metric. Basically, we know that the geometry side of the problem 
is obtained from the Einstein-Hilbert action, expressed in terms of the Ricci scalar $R$,  
\be
I_\mathrm{EH} = \int R \sqrt{-g}\ d^4x  \ .
\ee
Following the standard procedure, this leads to
\be
\delta I_\mathrm{EH} =  \int G_{ab}\delta g^{ab}\sqrt{-g}\ d^4x = \int \left( R_{ab} -{1\over 2} g_{ab}R \right) \delta g^{ab}\sqrt{-g}\ d^4x \ , 
\ee
where $G_{ab}$ is the Einstein tensor. Now, in the coupled  matter-gravity system
we have
\be
I = I_\mathrm{EH} + I_\mathrm{M} = \int \left( {1\over 2\kappa} R + \Lambda \right)\sqrt{-g}\ d^4x \ , 
\ee
where the coupling constant $\kappa(=8\pi G/c^4)$ is determined from the correspondence with Newtonian gravity in the appropriate limit.
This system leads to the usual Einstein field equations
\begin{equation}
    G_{ab} = \kappa T_{ab} \ ,  
\end{equation}
provided that 
\begin{equation} 
     T_{ab} = - \frac{2}{\sqrt{- g}} 
    { \delta \left(\sqrt{- g} \Lambda \right) \over  \delta g^{ab}}
  \ ,
  \ee
  or, equivalently,
  \be
   T^{ab}= \frac{2}{\sqrt{- g}} {\delta 
     \left(\sqrt{- g} \Lambda \right) \over  \delta g_{ab}}
     \ . \label{seten} 
\end{equation} 

Returning to the fluid problem, we see from \eqref{vars}  that the multi-fluid stress-energy tensor takes the form
\be
     T^{ab}_\mathrm{M} = \Psi g^{ab}+ \sum_{\X} 
                       n^a_\X \mu_\X^b \ . \label{seten2} 
\ee
It is worth noting that 
when the set of fluid equations, \eqref{consv2} and \eqref{fas}, is satisfied then it is 
automatically true that $\nabla_{a} T^{ab}_\mathrm{M} = 0$. 

Provided we are given the appropriate matter Lagrangian (a far from trivial problem as a realistic model should build on microphysics including the various relevant interactions) we now have all the equations we need
to describe the dynamics of the fluid system,  its effect on the gravitational field and vice versa.

As an aside, it is worth noting that the variational model is more general than the typical multi-component models considered in the literature (especially in cosmology) 
as they tend to assume the existence of partial pressures (see \cite{mark} for a relevant discussion in the present context). 

\subsection{Electromagnetism}

Let us now consider electromagnetism in Einstein's theory. As usual \cite{efsth}, we 
construct the relativistic version of Maxwell's equations by means of a variational argument with respect to the vector potential $A^a$.
The corresponding Lagrangian is built from the anti-symmetric Faraday tensor;
\be
F_{ab} = 2 \nabla_{[a} A_{b]}  \ .
\ee 
We also need to couple the electromagnetic field to the matter flow, represented by the charge current $j^a$. Letting the relevant coupling constant be 
$\mu_0$, the action takes the form~\footnote{Throughout the discussion we assume that the matter is electromagnetically passive in the sense that its properties are not effected by the presence of the electromagnetic field. It would be relatively easy to generalise the description to account for polarisation effects etcerera, but this results in a model that is slightly less transparent which may detract attention from the key issues that we want to focus on. We will address the general problem elsewhere.}
\be
I_\mathrm{EM}= \int L_\mathrm{EM}\sqrt{-g}\ d^4 x \ , 
\ee
with
\be
L_\mathrm{EM} = - {1 \over 4 \mu_0}F_{ab}F^{ab} + j^a A_a \ .
\label{lem}\ee
However, the  current term in this expression is not gauge-invariant. Under a gauge transformation of the vector potential, i.e. exercising the freedom to add the gradient of an arbitrary scalar field $\psi$, the second term in \eqref{lem}   transforms as
\be
j^a A_a \to j^a A_a + j^a \nabla_a \psi = j^a A_a + \nabla_a \left( \psi j^a\right) - \psi \left( \nabla_a j^a\right) \ .
\ee
The second term on the right-hand side will contribute a surface term to the action integral, and hence can be ``ignored'' in the usual way.
The third term is different. In order to ensure that the action is gauge-invariant, we must demand that the current is 
conserved, i.e. that
\be
\nabla_a j^a = 0  \ .
\ee
The field equations that we derive require that this constraint be satisfied.

With an action in hand it is straightforward to work out the variation with respect to the vector potential (keeping $j^a$ fixed!), and we arrive at the standard result;
\be
\nabla_b F^{ab} = \mu_0 j^a \ .
\ee
 The relativistic Maxwell equations are completed by 
\be
\nabla_{[a}F_{bc]}=0 \ ,
\ee
which is automatically satisfied given the anti-symmetry of  $F_{ab}$.

Finally, a variation with respect to the metric leads to the electromagnetic stress-energy tensor being given by
\be
T_{ab}^\mathrm{EM} =  {1\over \mu_0} \left[g^{cd}F_{ac}F_{bd}-{1\over 4}g_{ab}\left( F_{cd}F^{cd}\right) \right] \ .
\label{TEM}\ee
It is worth noting that this leads to
\be
\nabla_a T^{ab}_\mathrm{EM} =  j_a F^{ab} \equiv - f_\mathrm{L}^b \ ,
\label{loren}\ee
which (as we will see later) defines the Lorentz force $f_\mathrm{L}^a$.

In principle, the electromagnetic dynamics is now fully specified, as we can solve the system for the vector potential $A^a$. 
However, in most applications it is more intuitive to work with the electric and magnetic fields $E^a$ and $B^a$. The down-side to this is that these are 
observer dependent quantities. 
This is obvious since varying electric fields generate magnetic fields and vice versa, and the induced variation depends on the motion of the observer. 

According to an observer moving with four-velocity $u^a$, the Faraday tensor can be expressed as~\footnote{Electromagnetism is complicated by the fact that there are different conventions regarding units, signs etcetera. Our discussion, that follows \cite{carter79},  differs from alternatives like \cite{ellis} in a few subtle ways.  First of all the sign of the magnetic field $B^a$ is different, but this is later compensated for by a difference in the definition of $\epsilon_{abc}$ which is used to represent the spatial curl. These differences mean that any comparison with the literature must be carried out with care. The model we develop is completely self-consistent and natural in that it leads to the anticipated weak-field, low velocity results.}
\be
F_{ab} =  2 u_{(a} E_{b)}  + \epsilon_{abcd}u^c B^d \ , 
\label{Faraday}\ee
(where round brackets indicate symmetrization).
This defines the electric and magnetic fields as
\be
E_a = -  u^b F_{ba} \ , 
\ee
and
\be
B_a = - u^b \left( {1 \over 2} \epsilon_{abcd}F^{cd}\right) \ .
\ee
The physical fields are both orthogonal to $u^a$, so 
each field has three components, just as in non-relativistic physics. We also need an expression for the current, and it is natural 
to decompose this in a similar way;
\be
j^a = \sigma u^a + J^a \ , \qquad \mbox{where} \qquad J^a u_a = 0  \ .
\ee

\subsection{A  comfortable marriage}

So far, we have done quite a lot of preparatory work, going over standard territory without adding any real new insight. Our patience
 with this exercise is about to pay off, as we will now be able to make swift progress. This illustrates the 
  advantage of having a well-grounded action principle for coupled fluids, and an 
identification of the true momenta, and shows how easy it is to incorporate
electromagnetism into the multi-fluid system \cite{samuel}.  We simply  need to consider multiple charge carriers with identifiable fluxes, $n_\x^a$, and 
individual charges, $q_\x$, such that the charge current associated with each flow is 
\beq
    j^a_\x = q_\x n^a_\x \ ,
\eeq
and the total current, that sources the electromagnetic field, is given by the sum
\be
j^a = \sum_\x j_\x^a \ .
\ee
It is worth recalling that the variational derivation in Section~IIB requires that the current is conserved. However, this constraint 
is \underline{automatically} satisfied if each individual current is conserved, as  assumed in the variational multi-fluid model. Hence, we simply have to change the  
electromagnetic Lagrangian to 
\beq
  L_\mathrm{EM} = - \frac{1}{4\mu_0} F_{ab} F^{ab} + 
                           A_a  \sum_\x j^a_\x \ , 
\eeq
to combine the two models.

It is easy to see that the equations that govern the electromagnetic field remain exactly as before. However,
the  coupling to the current leads to modified fluid momenta; 
\beq
    \bar{\mu}^\x_a = \mu^\x_a + q^\x A_a \ , \label{guagecoup}
\eeq
which satisfy the equations of motion
\beq
    2n^a_\x \nabla_{[a} \bar{\mu}^\x_{b]}  = 0 \ .
\label{mageul}\eeq
As an alternative, we can write this as an explicit  force-balance relation.
Moving the electromagnetic contribution to the right-hand side, we get 
\be
f^\x_a =  2 n_\x^b \nabla_{[b} \mu^\x_{a]} 
= q^\x n_\x^b  F_{ab} = j_\x^b F_{ab}  \ .
\label{forces}\ee
To see that this result makes sense, note that the total 
energy-momentum tensor is easily obtained as the sum of the two previous expressions;
\beq
  T^{ab}=T^{ab}_\mathrm{M}+T^{ab}_\mathrm{EM} \ .
\eeq
This means that we must  have
\be
\nabla_a T^{ab}_\mathrm{M} = -  \nabla_a T^{ab}_\mathrm{EM} = - j_a F^{ab} = f_\mathrm{L}^b \ .
\ee
In other words, the combined system is such that
\be
f^\mathrm{L}_b = \sum_\x f^\x_b  \ .
\ee

The variational formalism naturally lends itself to a consideration of conserved quantities, like the magnetic helicity \cite{carter}. The
discussion becomes particularly elegant if carried out using the language of differential forms~\footnote{We are not making explicit use of differential geometry and forms in this paper, as we want the results to be directly useful to a more general audience, e.g. in astrophysics.}. We will not discuss 
conservation laws in this paper, but the interested reader can find relevant recent discussions in \cite{gour1,gour2}.

Before we proceed, it is worth digressing on the fact that the charge current does not enter the electromagnetic stress-energy tensor \eqref{TEM}. As this is a key (albeit somewhat technical) point, it is worth demonstrating the result in detail. To do this, let us focus on the contribution to the total action from the matter-field coupling;
\be
I_\mathrm{C} = \int j^a A_a \sqrt{-g}\ d^4 x \ .
\ee 
Variation of the integrand then leads to a sum of terms of the form
\be
\delta \left( n_\X^a A_a \sqrt{-g}\right)  = \sqrt{-g} \left[ A_a \delta n_\X^a + n_\X^a \delta A_a \right] +n_\X^a A_a \delta  \sqrt{-g}  \ .
\label{var1}\ee
Naively, the first term affects the Euler equation, the second leads to the current term in the Maxwell equations and the final term 
should enter the stress energy tensor.
However, the last contribution is cancelled by a term originating from the variation of the matter flux. 
Using \eqref{dna}  in \eqref{var1} (ignoring surface terms) we arrive at 
\be
\delta \left( n_\X^a A_a \sqrt{-g}\right)  = \sqrt{-g} \left(2 \xi_\X^a n^b \nabla_{[a} A_{b]} + n_\X^a \delta A_a \right) \ .
\label{var2}\ee
The first term enters the Euler equations \eqref{forces} and the second leads to the current term in the Maxwell equations. The electromagnetic contribution to the stress-energy tensor is completely determined by the first term in \eqref{lem}, leading to \eqref{TEM}.

\section{Resistive magnetohydrodynamics}

The formalism developed in the previous section provides a general framework for describing the dynamics of charged multi-fluid systems
in relativity. However, as the model arises from a variational analysis it does not account for dissipative mechanisms. Hence, we need to amend it
if we want to model, for example, resistivity. This is obviously a key aspect if we want to be able to model the evolution of electromagnetic 
fields in various astrophysical and cosmological settings. However, we  know that the general dissipative problem is a severe challenge in relativity. 
We also know that many different dissipation channels may affect a generic multi-fluid system \cite{extro,monster}. Hence, we set a more modest target and explore the role of a simple, phenomenological, resistivity. As it turns out, the problem involves tricky issues already at this level.  In general, any dissipative mechanism will generate heat, so a realistic model must account for the associated heat flux. However, this problem is known to be associated with both causality and stability issues in relativity \cite{is1,is2,hislin1,hislin2}. These problems can be resolved \cite{cesar1,cesar2}, but we must proceed carefully. 

Given the various issues involved, we consider the resistive problem at two different levels of sophistication. First (in this section) we consider a cold plasma, where the various relative velocities are sufficiently low that the problem simplifies. Having understood this problem  we proceed (in the next section) to consider the general problem, 
with arbitrary velocities and the presence of a heat flow. In each case, we consider a system with two charge carriers,  with  
 individual particles carrying a single unit of charge. This means that the models apply to both pair-plasmas with positrons and electrons and proton-electron plasmas. 
These examples provide useful illustrations and the discussion serves to highlight differences between the two systems.

Throughout the discussion, we  assume that the system is fully ionized. 
That is, we do not allow for the presence of a charge neutral component, as would be required if we wanted to model a magnetized neutron star core, for example. 
The inclusion of such a component is, in principle, straightforward although the algebra obviously gets more involved (especially if one accounts for entrainment).

\subsection{Choice of frame}

We focus on a two-component system with one component, labelled p, carrying a single positive unit of charge $q_\p=e$ while the 
other component, labelled e, carries a single negative unit of charge $q_\e=-e$. The associated charge currents are
$j_\p^a=en_\p^a$ and $j_\e^a=-en_\e^a$, respectively. We will not, initially, make any assumptions regarding the relative masses of the two components.
This means that the model applies to both pair plasmas and proton-electron systems (indeed, any  two-component system with electrons and single charged ions). 

A key aim of the exercise we are embarking on is to derive the relativistic version of Ohm's law. Basically, we want to start from the  charged
two-fluid system and arrive at a model from which the assumptions associated with standard relativistic magnetohydrodynamics become clear.
This discussion will obviously involve both the electric and the magnetic field, as well as the charge currents. Now, we know that $E^a$ and $B^a$ are 
observer dependent quantities. Hence, the model  involves a judicious choice of observer. It is natural to begin by considering this issue.

Given an observer with four velocity $u^a$ (normalised such that $u^au_a=-1$) we can decompose the two fluxes;
\be
n_\x^a = n_\x u_\x^a \ , \quad \mbox{where} \qquad n_\x^2 = - n_\x^a n^\x_a \qquad \mbox{and} \qquad u_\x^a u^\x_a = - 1 \ ,
\ee
using
\be
u_\x^a = \gamma_\x \left(u^a + v_\x^a\right) \ , \qquad \mbox{where} \qquad u^a v^\x_a = 0 \ , \qquad \mbox{and} \qquad \gamma_\x = \left( 1 - v_\x^2 \right)^{-1/2} \ .
\ee
In the first instance, we will assume that the ``drift'' velocities $v_\x^a$ are small enough that we can linearise the model, i.e. assume that 
$\gamma_\x\approx 1$. This model should be relevant for cold plasmas~\footnote{It it worth pointing out that we do not impose any restrictions on the bulk velocity in the final single-fluid model here. The fluid, as represented by $u^a$, may move fast relative to an inertial observer. It is the relative velocity $v_\x^a$ with respect to $u^a$ that is assumed to be small.}

We also need the fluid momenta, which would generally involve entrainment between the two components. However, as we are not aware of a
physical argument for the presence of entrainment between protons and electrons (or, indeed, positrons and electrons), we do not account for this effect here (although we will consider it when we discuss heat flux and entropy later). 
This means that we have 
\be
\mu^\x_a = \B^\x n^\x_a= \B^\x \gamma_\x n^\x  \left( u_a
+ v^\x_a \right) \ .
\ee
The chemical potential of each component is generally defined by 
\be
\mu_\x = - u_\x^a \mu^\x_a =  n_\x \B^\x \ ,
\ee
which means that, 
\be
\mu^\x_a = \mu_\x \gamma_\x  \left( u_a
+ v^\x_a \right) \approx \mu_\x  \left( u_a
+ v^\x_a \right) \ .
\ee

With these definitions we can write the (linearised) fluid stress-energy tensor as
\be
T^\mathrm{M}_{ab} = \Psi g_{ab} + \left( n_\p \mu_\p + n_\e \mu_\e \right) u_a u_b +\left( n_\p \mu_\p v^\p_b + n_\e \mu_\e v^\e_b \right) u_a
+ \left( n_\p \mu_\p v^\p_a + n_\e \mu_\e v^\e_a \right) u_b \ .
\ee
Contracting this with $u^a$ we get an expression for the momentum flux;
\be
u^a T^\mathrm{M}_{ab} = \left( \Psi -n_\p \mu_\p - n_\e \mu_\e \right) u_b  -  \left( n_\p \mu_\p v^\p_b + n_\e \mu_\e v^\e_b \right) \ .
\label{moms}\ee
Contracting with  $u^b$ again, we find that the energy density measured by the observer is
\be
\rho = u^a u^b T^\mathrm{M}_{ab} = - \Psi+ n_\p \mu_\p + n_\e \mu_\e  \ .
\ee
We see that, in the linear model $\Psi$ is the pressure. Hence, we replace it with $P$ in the following, leading to the anticipated 
thermodynamic relation (the integrated first law)
\be
P+\rho =  n_\p \mu_\p + n_\e \mu_\e \ .
\label{0law}\ee
Note that, as we are only considering the fluid contribution here, our definitions of  $P$ and $\rho$ do not include electromagnetic effects (i.e. the magnetic pressure is not accounted for yet). 

From \eqref{moms} we see that we can choose observers such that there is no relative (fluid) momentum flux by setting~\footnote{This ``centre-of-mass'' frame provides the natural relativistic extension of the standard plasma physics analysis \cite{schnack,bellan}. A common alternative strategy \cite{tsagas} is to use a frame in which there is no net particle flux, i.e. where
$$
n_\p v_\p^a + n_\e v_\e^a = 0\ .
$$
The final results should be the same in the two case, but as is evidenced by the relatively straightforward route to the final result, our chosen strategy is the most natural. It is also readily extended to the ``hot'' case, where heat flux is added to the problem, as in Section~IV.
}
\be
n_\p \mu_\p v^\p_b + n_\e \mu_\e v^\e_b = 0 \ .
\ee
This leads us to define a velocity $v^a$ such that
\be
\left(P+\rho\right) v^a = n_\p \mu_\p v_\p^a + n_\e \mu_\e v_\e^a  \ , 
\label{vdef}\ee
and highlights the relevance of the frame in which $v^a=0$. We express the second degree of freedom in terms of the relative velocity 
\be
w^a = v_\p^a - v_\e^a \ .
\ee
With these definitions we have
\be
v_\p^a = v^a + {n_\e \mu_\e \over P+\rho} w^a\ ,
\ee
and
\be
v_\e^a = v^a - {n_\p \mu_\p \over P+\rho} w^a\ ,
\ee
and the charge current  takes the form
\be
j^a  = e \left(n_\p-n_\e\right) \left(u^a+v^a\right) + e{n_\p n_\e \over P+\rho} \left( \mu_\p + \mu_\e\right) w^a  \ .
\ee
From this result we read off the charge density $\sigma = e \left(n_\p-n_\e\right)$ in the observer's frame.
If we assume that the system is charge neutral on macroscopic scales, a natural assertion for systems where the charge carriers (like the electrons) are 
highly mobile and one of the key assumption in standard magnetohydrodynamics, then the current simplifies to 
\be
j^a = J^a = e{n_\p n_\e \over P+\rho} \left( \mu_\p + \mu_\e\right) w^a  \ .
\ee

Moreover, in the case of a charge neutral plasma we have $P+\rho = n_\e \left( \mu_\p + \mu_\e\right)$ which means that the current takes the final form 
\be
J^a =  e n_\e  w^a \ .
\label{cure}\ee

\subsection{The resistivity}

In order to account for the resistivity, we need to add a phenomenological ``force'' term to \eqref{forces}.
This additional term should represent the dissipative interaction between the two components, and from
non-relativistic intuition \cite{schnack,bellan}, we expect it to be linear in the relative velocity between the two components. We also see from \eqref{forces} that the 
required force must be orthogonal to each respective flux (note that this condition must be relaxed if we want to allow for particle creation/destruction). Based on these points, we let the resistive forces take the form
\be
\tilde f_\p^a = e \R \perp_\p^{ab} n^\e_b = -  \R \perp_\p^{ab} j_b \ , 
\ee
and
\be
\tilde f_\e^a = e \R \perp_\e^{ab} n^\p_b =   \R \perp_\e^{ab} j_b\ .
\ee
These expressions represent linear scattering of the two components.  The
resistivity experienced by one component is proportional to the number of particles of the other kind that flows relative to it.

The resistivity is further constrained by the fact that the sum of the forces must vanish (essentially Newton's third law). 
This follows immediately from the fact that the divergence of
the  non-dissipative stress-energy tensor [which arises from the sum of \eqref{forces}] must vanish. 
With the suggested forces, we have
\be
\tilde f_\p^a + \tilde f_\e^a = e\R \left( \perp_\p^{ab}n^\e_b + \perp^{ab}_\e n^\p_b\right) \approx e \R \left( n_\p - n_\e \right) w^a \ .
\label{fcomb}\ee
This shows that the linearized model is only consistent as long as the system is charge neutral. If there is charge imbalance, we need to alter the model.
At first sight, this may seem surprising but it is actually quite natural. The model only accounts for the two charged components, whereas
the general system would also have the heat generated by the dissipation. The correct interpretation of \eqref{fcomb} is that, for a charge neutral system, there is no heat generated at the linear level.  In order to consider a more general system, we need to account  for the heat. Then the force balance is ensured by introducing an additional component, which we will take to be the entropy, with a 
corresponding force of the required form. We will discuss this extended system in the next section. For now, we simply assume charge neutrality and note that 
the corresponding low-velocity model describes a ``cold plasma'' in the sense that there is no heat generated in the system.

\subsection{Generalized Ohm's law}

The problem under consideration has two fluid degrees of freedom, represented by \eqref{forces} with the added resistivity terms (on the right-hand side). 
One can (obviously) combine these two equations in different ways. It seems natural to adapt the standard strategy
from non-relativistic plasma physics \cite{schnack,bellan} and consider a ``total momentum'' equation alongside a suitably weighted difference. The first of these 
equations  follows by adding \eqref{forces}, and from the discussion in the previous section we know that this leads to 
\be
\nabla_a T^{ab} = 0 \ ,
\label{divT}\ee
as the sum of the resistive forces vanishes (to linear order). In order to represent the second degree of freedom, we divide the two equations from \eqref{forces} by 
$n_\x \mu_\x$ and then take the difference. The weighting (different from that used in other recent discussions of the problem \cite{tsagas}) is motivated by the Newtonian limit, where $\mu_\x \to m_\x$ (the rest mass) and corresponds to the "centre-of-mass" frame. With this weighting the difference equation simplifies considerably. One may obtain the same final result with a different weighting, but the analysis would then have to make explicit use of the total momentum equation \eqref{divT} in simplifying the expressions. Our route is more direct.

The difference equation that we require is made up of three pieces.
Considering first the fluid contribution to \eqref{forces} and the  definition of the chemical potentials, we have
\be
2 n_\x^a \nabla_{[a} \mu^\x_{b]}  = n_\x \perp^a_{\x b} \nabla_a \mu_\x + n_\x \mu_\x u_\x^a \nabla_a u^\x_b \ .
\ee
We also have 
\be
\perp_\x^{ab} = g^{ab} +  u_\x^a u_\x^b \approx \perp^{ab} + 2 u^{(a} v_\x^{b)} \ , 
\ee
where $\perp^{ab}=g^{ab}+u^a u^b$ is the projection orthogonal to $u^a$. Using these results, we find that the weighted difference (let us call it $f^\mathrm{D}_a$) in the linear model takes the form;
\begin{multline}
f^\mathrm{D}_b={2 \over n_\p\mu_\p}  n_\p^a\nabla_{[a} \mu^\p_{b]}  - 
{2 \over n_\e \mu_\e} n_\e^a  \nabla_{[a} \mu^\e_{b]}    \\
= u_\p^a\nabla_a u^\p_b - u_\e^a\nabla_au^\e_b + {1\over \mu_\p} \perp_{\p b}^a \nabla_a \mu_\p
-{1\over \mu_\e} \perp_{\e b}^a \nabla_a \mu_\e \\
\approx
u^a \nabla_a w_b  + w^a \nabla_a u_b + {1\over \mu_\p} \perp_{\p b}^a \nabla_a \mu_\p
-{1\over \mu_\e} \perp_{\e b}^a \nabla_a \mu_\e \ .
\end{multline}
The last two terms expand to
\begin{multline}
{1\over \mu_\p} \perp_{\p b}^a \nabla_a \mu_\p
-{1\over \mu_\e} \perp_{\e b}^a \nabla_a \mu_\e \\
\approx
\left(\perp^a_{\ b} + u^a v_b + u_b v^a \right) \left( {1\over \mu_\p}  \nabla_a \mu_\p
-{1\over \mu_\e}\nabla_a \mu_\e \right) \\
+ {1 \over (P+\rho) \mu_\p \mu_\e} \left(u^a w_b + u_b w^a \right) \left[ {n_\e \mu_\e^2 \nabla_a \mu_\p + n_\p \mu_\p^2 \nabla_a \mu_\e} \right] \ .
\end{multline}
This expression obviously simplifies somewhat in the frame where $v^a=0$. The result also simplifies for the two specific examples we are considering. 
For pair plasmas we have $\mu_\p=\mu_\e\equiv\mu$ so in the chosen frame the final result would be 
\be
f^\mathrm{D}_b= u^a \nabla_a w_b  + w^a \nabla_a u_b + {2\over \mu}  u_{(a} w_{b)}  \nabla^a \mu \ .
\ee
Meanwhile, for a proton-electron plasma we may assume that $\mu_\e \ll \mu_\p$ in which case (when $v^a=0$) we are left with~\footnote{It is worth being a little bit more precise here. In general, one can extract the rest-mass contribution ($m_\x$) to the chemical potential to get 
$$
\mu_\x = m_\x c^2 + \mu^\mathrm{N}_\x \ .
$$
At low velocities, where the second term can be neglected, it is obviously the case that $\mu_\e \ll \mu_\p $ for proton-electron plasmas. 
It is also clear that we have no reasons to assume that the gradients
$$
\nabla_a \mu_\x = \nabla_a \mu^\N_\x \ ,
$$
obey a similar ordering.
}
\be
f^\mathrm{D}_b\approx u^a \nabla_a w_b  + w^a \nabla_a u_b + 
\perp^a_{\ b}   \left( {1\over \mu_\p}  \nabla_a \mu_\p
-{1\over \mu_\e}\nabla_a \mu_\e \right) \\
+{2 \over \mu_\e}    u_{(a} w_{b)} \nabla^a \mu_\e \ .
\ee
In each case we can replace the relative velocity $w^a$ with the charge current via Eq.~\eqref{cure}. 

The expressions we have obtained represent the left-hand side of the equation that we are developing. The right-hand side is made up of two pieces. The first is the weighted difference of the two magnetic forces from \eqref{forces}. That is, we have 
\begin{multline}
f^\mathrm{M}_b = {e \over n_\p \mu_\p} n_\p^a F_{ba} +  {e \over n_\e \mu_\e} n_\e^a F_{ba} \\
=  e \left({\mu_\p + \mu_\e \over \mu_\p\mu_\e}\right) E_b + e \left( {1 \over \mu_\p} v_\p^a +  {1 \over \mu_\e} v_\e^a\right) F_{ba} \\
=
e \left({\mu_\p + \mu_\e \over \mu_\p\mu_\e}\right) \left( E_b + v^a F_{ba} \right) -  { e \left( n_\p \mu_\p^2 - n_\e \mu_\e^2 \right) \over (P+\rho) \mu_\p \mu_\e } 
w^a F_{ba}  \ .
\end{multline}
As before, this simplifies in the frame where $v^a=0$. In addition, for (charge neutral) pair plasmas the second term vanishes identically and we are left with  
\be
f^\mathrm{M}_b =e {\mu_\p + \mu_\e \over \mu_\p\mu_\e}  E_b  = {2e \over \mu} E_b \ .
\ee
Meanwhile, for a proton-electron system we would have
\be
f^\mathrm{M}_b \approx {e \over \mu_\e} \left(  E_b - 
w^a F_{ba} \right) \ .
\ee

In this case we need to consider the remaining term involving the Faraday tensor in more detail. From \eqref{Faraday} it is easy to see that we will have
\be
w^a F_{ba} = u_b \left(w^aE_a\right) + \epsilon_{bcd}w^c B^d \ ,
\ee
where we have defined $\epsilon_{bcd} = \epsilon_{abcd}u^a$ in order to make the final term resemble the standard three-dimensional cross product~\footnote{Note different conventions here, with some work defining $\epsilon_{bcd} = \epsilon_{bcda}u^a$ together with the opposite sign of the magnetic field, see e.g. \cite{ellis}.}. 
Since $w^a$ is proportional to the charge current, we recognize the two terms as the Joule heating and the Hall effect, respectively. These effects are notably absent in the linear model for a pair plasma \cite{koide08,koide09,tsagas}.

Finally, we need the weighted difference between the two resistivities. That is;
\begin{multline}
f^\mathrm{R}_b  = {1 \over n_\p \mu_\p}  \tilde f^\p_b - {1 \over n_\e \mu_\e}  \tilde f^\e_b = {e\R \over n_\p \mu_\p}\perp^\p_{ab} n_\e^a
-   {e\R \over n_\e \mu_\e}  \perp^{\e}_{ab} n_\p^a \\
= - \R \left(  {1 \over n_\p \mu_\p}\perp^\p_{ab} 
+   {1 \over n_\e \mu_\e} \perp^\e_{ab} \right) j^a \approx - \R {P+\rho \over (n_\p \mu_\p) (n_\e \mu_\e) } J_b \ ,
\end{multline} 
where the fact that the current is proportional to $w^a$ in the charge-neutral case has allowed us to neglect the various $v_\x^a$ terms in the 
projections.

The final result now follows from the combination
\be
f^\mathrm{D}_a = f^\mathrm{M}_a + f^\mathrm{R}_a   \ .
\label{fbal}\ee
In the general charge neutral case we have, in the frame where $v^a=0$,
\begin{multline}
e {\left(P+\rho\right) \over n_\e \mu_\p\mu_\e} E_b -  { e \left( \mu_\p - \mu_\e \right) \over \mu_\p \mu_\e } \left[
u_b \left(w^aE_a\right) + \epsilon_{bcd}w^c B^d\right]  - \R {P+\rho \over (n_\p \mu_\p) (n_\e \mu_\e) }J_b \\
=
u^a \nabla_a w_b  + w^a \nabla_a u_b  + \perp^a_{\ b}   \left( {1\over \mu_\p}  \nabla_a \mu_\p
-{1\over \mu_\e}\nabla_a \mu_\e \right) \\
+ {2 \over (P+\rho) \mu_\p \mu_\e} u_{(a} w_{b)}  \left[ {n_\e \mu_\e^2 \nabla^a \mu_\p + n_\p \mu_\p^2 \nabla^a \mu_\e} \right] \ .
\end{multline}

This result can be simplified by projecting out the contribution along $u^a$ (which will not affect  the expression for the electric field). This leads to the 
``final'' result;
\begin{multline}
e {\left(P+\rho\right) \over n_\e \mu_\p\mu_\e} E_b -  { e  \left(  \mu_\p -  \mu_\e \right) \over \mu_\p \mu_\e }  \epsilon_{bcd}w^c B^d  - \R {P+\rho \over (n_\p \mu_\p) (n_\e \mu_\e) } J_b \\
= \perp^{a}_{\ b} \left[ u^c \nabla_c w_a + {1\over \mu_\p}  \nabla_a \mu_\p
-{1\over \mu_\e}\nabla_a \mu_\e \right] +  w^a \nabla_a u_b  \\
+ {n_\e \over (P+\rho) \mu_\p \mu_\e} w_b u^a \left[ { \mu_\e^2 \nabla_a \mu_\p + \mu_\p^2 \nabla_a \mu_\e} \right] \ .
\label{genohm}\end{multline}

In these expressions, it may be useful to decompose $\nabla_a u_b$ in the standard way, see for example \cite{ellis}. That is, we 
use
\be
\nabla_a u_b  = \sigma_{ab} + \omega_{ab} - u_a \dot{u}_b + {1 \over 3} \theta \perp_{ab}  \ ,
\label{decomp}\ee
in terms of the expansion scalar
\be
\theta = \nabla_a u^a \ ,
\ee
the shear
\be
 \sigma_{ab} = D_{\langle a}u_{b\rangle} \ , 
\ee
where the angle brackets indicate symmetrization and trace removal, and
\be
D_a u_b = \perp_a^{\ c} \perp_{b}^{\ d} \nabla_c u_d \ .
\ee
The merit of using this (totally projected) derivative is that the individual terms in \eqref{decomp} are perpendicular to $u^a$.
 We have also defined the vorticity~\footnote{Note that we define the vorticity tensor to have the opposite sign compared to \cite{ellis}. This is obviously just convention, but it is important to keep it in mind if one wants to compare the various final relations.}
 \be
 \omega_{ab} = D_{ [a}u_{b]} \ .
 \ee
 The decomposition \eqref{decomp} makes the coupling between the charge current $J^a$ and the nature of the fluid motion more explicit.

The final relation for  pair plasmas can now be written~\footnote{ We have neglected a term
$$
 \left({\dot\mu/ \mu}\right) J_b \approx  \left({\dot\mu_\e^\mathrm{N}/ m_\e c^2}\right) J_b  \ ,
$$
in the bracket on the right-hand side. This should be a valid approximation at low velocities. We have also used
$$
\left( {\dot{n}_\e / n_\e } \right) J_b  \approx - \theta_b \ ,
$$
which  is true at the linear (in the current)  level since $\nabla_a  n_\e^a = 0$.
}
\be
E_b - {\R \over  en_\e} J_b = {\mu \over 2e^2 n_\e} \left[ \perp_{ab} \dot{J}^a + J^a \left(  \sigma_{ab} + \omega_{ab} + {4 \over 3} \theta \perp_{ab} \right)   \right] \ , 
\ee
where the dot represents the comoving time derivative $u^c \nabla_c$. As already mentioned, this expression is notable for the absence of the Hall effect, i.e. there is no term proportional to $\epsilon_{abc}J^b B^c$. 

The case of a proton-electron plasma is only slightly  more complicated. After neglecting $\mu_\e$ compared to $\mu_\p$, we end up with
\be
E_b -{1\over e n_\e} \epsilon_{bcd}J^c B^d - {\R \over e n_\e} J_b = {\mu_\e \over e^2 n_\e} \left[ \perp_{ab} \dot{J}^a + J^a  \left(  \sigma_{ab} + \omega_{ab} + {4 \over 3} \theta \perp_{ab} \right)   \right] - {1 \over e} \perp^a_{\ b} \nabla_a \mu_\e \ .
\ee
In this case, the Hall term is obviously present. We also have a ``Biermann battery'' term, $\perp^a_{\ b} \nabla_a \mu_\e$, which would serve to generate a magnetic field even if there was no field initially \cite{tsagas}. 

It is easy to show that our final results agree perfectly with the results obtained in \cite{tsagas}.

Before moving on, it is useful to consider the relation between our results and the common starting point for discussions of resistive effects in numerical simulations \cite{wata,pale,taka}. Much of the relevant literature builds on the work by Bekenstein and Oron \cite{bekor}.
Ignoring the right-hand side of Ohm's law in  the proton-electron case we have
\be
E_b ={1\over e n_\e} \epsilon_{bcd}J^c B^d + {\R \over e n_\e} J_b ={\R \over n_e e} \left( \perp_{ab} + {1\over \R} \epsilon_{bacd} u^c B^d \right) J^a
= S_{ba}J^a \ .
\label{geom}\ee
Define $\zeta = 1/\R$ and $\sigma = \R/n_\e e$ to get
\be
S_{ba} = {1 \over \sigma} \left( \perp_{ba} + \zeta \epsilon_{bacd} u^c B^d \right)  \ .
\ee
Inverting this, we arrive at 
\be
J_b = \sigma^{ab} E_b \ ,
\ee
with 
\be
\sigma^{ab} = {\sigma \over 1 +\zeta^2 B^2} \left( \perp^{ab} + \zeta^2 B^a B^b - \zeta \epsilon^{abcd} u_c B_d \right) \ .
\label{sig}\ee
This is the result stated in \cite{bekor}, once we account for the different sign conventions. At this point, we can make an important observation. 
It is easy to identify the Hall effect in the initial expression \eqref{geom}, but its presence is more convoluted in the alternative expression \eqref{sig}.
That this can lead to conceptual confusion is evidenced by \cite{zanotti}, where numerical evolutions for a truncated form of \eqref{sig} are carried out. The considered model includes a peculiarly amputated Hall effect, the actual meaning of which is unclear. This lesson tells us that an understanding of the physical origin of the model is imperative. 
 
\subsection{Towards ideal MHD}

We are now in a position where we can assess the relative importance of the different terms in the generalized version of Ohm's law \eqref{genohm}.
Let us first consider under what conditions we can neglect the inertia of the charge current compared to the resistivity. In order to do this we need
\be
{\R \over e n_\e} J \gg {\mu_\e \over e^2 n_\e} \dot{J} \approx {m_\e \over e^2 n_\e} \dot{J} \ .
\ee
It is natural \cite{bekor} to associate the resistivity with a relaxation timescale $\tau_r$ such that
\be
\R = {m_\e \over e\tau_r} \ .
\ee
If we also  assume that the the dynamics has a characteristic timescale $\tau_d$, such that $\dot J  \sim J/\tau_d$, then it is easy to see that we can neglect the inertia of the charge current as long as
\be
\tau_d \gg \tau_r\ .
\ee
When this condition holds, i.e. for sufficiently slow dynamics, the system is essentially oblivious of its plasma physics origins. This condition shows why ideal magnetohydrodynamics is a good model for slowly evolving, or stationary, systems.

Next let us compare the Hall term to the resistivity. The former dominates (in magnitude)  if 
\be
 B \gg  \R  \ , 
\ee
which, if we introduce the electron cyclotron frequency
\be
\omega_c = {e B \over m_\e} \ , 
\ee
leads to the condition
\be
\omega_c \tau_r \gg 1\ .
\ee
This means that the electron executes many cyclotron ``oscillations'' before the motion is damped. 

Finally, we need to establish when the resistivity can be neglected compared to the electric field. This requires
\be
E \gg {\R \over e n_\e} J = {m_\e \over n_\e e^2 \tau_r} J\ .
\ee
Here we need to make use of Maxwell's equations (see below), which lead to (in Gaussian units!)
\be
E \sim {V \over c} B\ ,
\ee
where we assume that the dynamics has a characteristic length-scale $L$, and an associated velocity $V=L/\tau_d$. This leads to the final condition
\be
J \ll n_e e \left( {V \over c} \right) \omega_c \tau_r \ll n_e e \left( {V \over c} \right)\ .
\ee
The last condition is required since we also want to be able to neglect the Hall term. We are essentially left with a low-velocity constraint. 

These rough estimates provide useful insight into the applicability of ``ideal'' magnetohydrodynamics, which corresponds to the assumption that $E^a\approx 0$. The usual argument for this is that the medium is a perfect conductor, i.e. $\R\to 0$. However, this limit only affects the resistive term in \eqref{genohm}. We still have to argue that the remaining terms are unimportant. This is not quite as easy. At the end of the day, ideal magnetohydrodynamics is more an assumption than an approximation \cite{schnack} (the interested reader may want to compare the present discussion to the variational derivation of magnetohydrodynamics in \cite{achter}).

\subsection{The remaining fluid equation}

So far, we have focused on the weighted difference between the two momentum equations in the plasma. To complete the 
``single-fluid'' model we need to express the remaining degree of freedom in terms of our chosen variables. It is natural to obtain the required
equation from \eqref{divT}. 

As a first step, we consider the non-magnetic contributions. As the individual number fluxes are conserved in the variational approach, we see that 
\begin{multline}
\nabla_a T_\mathrm{M}^{ab} = \nabla^b \Psi + u_\p^b n_\p^a \nabla_a \mu_\p  + u_\e^b n_\e^a \nabla_a \mu_\e +\mu_\p n_\p^a \nabla_a u_\p^b 
+ \mu_\e n_\e^a \nabla_a u_\e^b \\
\approx 
\perp^{ab}\nabla_a P + \left( P+\rho\right) \dot{u}^b +2 u^{(a}v_\p^{b)} n_\p \nabla_a \mu_\p
+2 u^{(a}v_\e^{b)} n_\p \nabla_a \mu_\e \\
+n_\p\mu_\p \left( \dot{v}^b_\p + v_\p^a\nabla_a u^b\right) +n_\e\mu_\e \left( \dot{v}^b_\e + v_\e^a\nabla_a u^b\right) \ .
\end{multline}
The first line is exact, while the second line holds at the level of linearised relative velocities. Expressing this result in terms of $v^a$ and $w^a$, we have
\begin{multline}
\nabla_a T_\mathrm{M}^{ab}\approx \left( P+\rho\right) \dot{u}^b + \perp^{ab}\nabla_a P + 2 u^{(a}v^{b)} \nabla_a \Psi \\ 
+
{n_\e^2 \mu_\p \mu_\e \over P+\rho} u^b w^a \left( 
{1\over \mu_\p} \nabla_a \mu_\p - {1\over \mu_\e}\nabla_a \mu_\e\right) 
+ \left(P+\rho\right)  \left( \dot{v}^b + v^a\nabla_a u^b\right) \ .
\end{multline}
Here we have used the fact that we are considering a charge neutral system.
This result provides the left-hand side of the final equation.
As discussed in section~IIC the matter contribution is balanced by the electromagnetic stresses, which provides the right-hand side for the equation we are interested in. This takes the form
\be
-\nabla_a T^{ab}_\mathrm{EM} = u^b \left(j_a E^a\right) + \epsilon^{bac} j_a B_c\ .
\ee 
The final equation will only have spatial components with respect to $u^a$, but the relations we have written down so far also have parallel components. 
However, if we contract the combined equation with $u_b$ and compare to what we get if we contract our generalized Ohm's law \eqref{genohm}
with the current, then we see that the two results agree (at the linearized level, of course). Hence, only the orthogonal component contains new information. In the frame where $v^a=0$ the final fluid equation takes the form
\be
\left( P+\rho\right) \dot{u}^b + \perp^{ab}\nabla_a P = \epsilon^{bac} J_a B_c\ .
\label{finmom}\ee
This is simply the perfect fluid equation of motion augmented by the Lorentz force. 

To complete the model, we may also consider the two conservation laws \eqref{consv2}. It is straightforward to show that the difference between these corresponds to the required conservation law for the charge current. Meanwhile, after making use of the component aligned with $u^a$ from the weighted difference equation that leads to \eqref{genohm} , the sum of the two conservation laws can be written
\be
J_a E^a = {P+\rho \over n_\e} \left( \dot{n}_\e + n_\e \theta \right)\ .
\ee
Moreover, one can show that (at the linear level) this expression also follows from the component of the total momentum equation \eqref{divT} that is aligned with $u^a$. 

At the end of the day we have two scalar equations and two equations governing velocity components that are spatial with respect to $u^a$. Thus, we have explicitly accounted for the degrees of freedom of the original two-fluid system. To complete the system we also need Maxwell's equations. Before 
discussing these, let us make a brief diversion and touch upon a model that is common for neutron star magnetospheres.

The conditions in the magnetosphere of a neutron star (or, indeed, a black hole) are expected to be such that there is sufficient plasma present to carry a charge current, but the associated inertia can be neglected \cite{msph,Thomps}. Thus, we may neglect the inertia in
\eqref{finmom}, essentially decoupling the matter from the magnetic problem. This leads to what is known as force-free electrodynamics \cite{uch,komm}. In these circumstances we would have 
\be
 \epsilon^{bac} J_a B_c \approx 0 \ ,
\ee
which implies that charges may only flow along the magnetic field. 

The force-free assumption can obviously be used independently of the assumptions that lead to ideal magnetohydrodynamics. Basically, one may envisage a range of different ``approximations'' depending on the circumstances. The force-free model simplifies magnetosphere modelling, but one must apply it with care since it breaks down near magnetic neutral points. In the context of the present discussion, it is also worth noting that \eqref{finmom} may be extended to include various dissipation channels (like shear viscosity) other than the pure collisional resistivity that we have accounted for. If the multi-fluid aspects are taken seriously \cite{formalism}, this may lead to a much more complex problem.

\subsection{Maxwell's equations}

Given an observer moving with $u^a$, representing a fibration of spacetime, the decomposition of Maxwell's equations is standard.  Nevertheless, we list the results here for completeness. For a more detailed discussion, see for example, \cite{ellis}.

First of all, 
\be
\nabla_a F^{ba} = \mu_0 j^b \ , 
\ee
leads to
\be
\perp^{ab} \nabla_b E_a = \nabla_a E^a - E^a \dot u_a = \mu_0 \sigma + \epsilon^{abc} \omega_{ab} B_c  = \mu_0 \sigma + 2W^a B_a\ ,
\ee
where we have defined the vorticity vector as
\be
W^{a} = {1 \over 2} \epsilon^{abc} \omega_{bc} \ , \quad \mbox{so that} \quad \omega_{ab} = \epsilon_{abc}W^c \ , \quad \mbox{and} 
\quad u^a W_a = 0 \ .
\ee
We also get
\be
\perp_{ab}\dot{E}^b - \epsilon_{abc}\nabla^b B^c + \mu_0 J_a =  \left( \sigma_{ab} -\omega_{ab} - {2\over 3} \theta \perp_{ab}\right) E^b + \epsilon_{abc}\dot{u}^b B^c \ .
\ee

Secondly, 
\be
\nabla_{[a}F_{bc]} = 0 \ ,
\ee
leads to
\be
\perp^{ab}\nabla_b B_a = - 2W^a E_a\ ,
\ee
and
\be
\perp_{ab}\dot{B}^b +  \epsilon_{abc}\nabla^b E^c   = - \epsilon_{abc}\dot{u}^b E^c +  \left( \sigma_{ab} -\omega_{ab} - {2\over 3} \theta \perp_{ab}\right) B^b \ .
\ee

It is easy to see that, if we consider an inertial observer,  these results reduce to the standard text-book form of Maxwell's equations. The complete  expressions given here are, of course, useful if we are interested in more general settings. In particular, they highlight the coupling between the electromagnetic field and a given fluid flow (with shear, vorticity and expansion). 

\section{Adding entropy: Hot plasmas}

The low-velocity model we have discussed so far is consistent and applicable to many situations of interest. It also provides a number of potentially important extensions of the ideal magnetohydrodynamics that tends to be used in relativistic astrophysics. However, as we have already hinted at, the model does not account for the presence of heat. This is an unfortunate omission since resistivity is a dissipative process and hence will be associated with  entropy variations constrained by the second law of thermodynamics. This effect turns out to be quadratic in the relative velocities, which is why we got away with neglecting it in the linear model. In a more general setting, we need to account for the induced heat flow. The problem of heat in relativity is, however, known to be thorny. A model needs to be constructed carefully in order to avoid unwanted  instabilities and causality violation \cite{is1,is2,hislin1,hislin2}. As recently demonstrated, one can construct a satisfactory model by treating the entropy as an additional fluid component~\footnote{The fluid entropy model is easily motivated by considering phonons in a system where the mean-free path for phonon-phonon scattering is suitably short compared to that associated with dissipative scattering off the material components.}, and accounting for entrainment between the entropy and the other components in the system \cite{cesar1,cesar2}. This entropy entrainment is closely associated with the inertia of heat and the finite thermal relaxation timescale that is required in order to avoid superluminal signal propagation. We develop our model with these key points in mind.

\subsection{Setting the stage: A three-fluid system} 

We consider a hot plasma consisting of the two charged components from the cold model, labelled p and e as before, and an additional entropy, which we label s. As the entropy plays a special role, being constrained by the second law, we will single out this component by letting its flux be given by $s^a=n_\s^a$ while the corresponding chemical potential is $ \Theta_a=\mu^\s_a$. The latter determines the temperature measured by a given observer. With these definitions we have the total stress-energy tensor \cite{livrev}
\be
T_{ab} = \Psi g_{ab} + n^\p_a \mu^\p_b + n^\e_a \mu^\e_b + s_a \Theta_b\ ,
\ee
where the generalized pressure $\Psi$ is defined as
\be
\Psi = \Lambda - n^a_\p \mu^\p_a - n^a_\e \mu^\e_e -s^a\Theta_a\ .
\ee
Following \cite{cesar1,cesar2} we 
account for entrainment between the entropy and each of  the material components (encoded in coefficients $\A^{\x\s}$). Thus,  we have the momenta
\be
\mu^\x_a = \B^\x n^\x_a + \A^{\x\s}s_a \ , \qquad \x=\p,\e\ ,
\ee
and
\be
\Theta_a = \B^s s_a + \A^{\p\s}n^\p_a + \A^{\e\s} n^\e_a\ .
\ee

As in the low-velocity model, we introduce a family of observers that allow us to define the electric and magnetic field components. We now have a number of different options. The strategy that we adopt provides a natural extension of the Eckart frame for a single component matter model, cf., \cite{cesar1,cesar2}. To be specific, we choose the observer frame to be such that the only relative momentum flow is due to the heat.

Defining first of all the number densities as measured in the respective fluid frames, we have~\footnote{In order to avoid the notation becoming unnecessarily cluttered we opt to distinguish quantities measured in the various rest-frames by hats. This means that the notation differs from that used in, say, \cite{livrev}, which is unfortunate. However, the main focus will be on quantities measured by the selected observer. As these quantities appear much more frequently than the rest-frame components, the chosen convention is more convenient.}
\be
\hat n_\x^2 = - n_\x^a n^\x_a  \ , \qquad \mbox{and} \qquad \hat s^2 = - s^a s_a\ .
\ee
Decomposing the velocities with respect to a specific observer moving with $u^a$ we then have
\be
n_\x^a = \hat n_\x \gamma_\x \left( u^a + v_\x^a \right) \ , \qquad u^a v^\x_a = 0 \ , \qquad \gamma_\x = \left( 1 - v_\x^2 \right)^{-1/2}\ , 
\ee
leading to the number density measured by the observer being given by
\be
n_\x =  - u_a n_\x^a = \hat n_\x \gamma_\x\ .
\ee
Similarly, we have 
\be
s^a = \hat s \gamma_\s \left( u^a + v_\s^a \right) \ , \qquad u^a v^\s_a = 0 \ , \qquad \gamma_\s = \left( 1 - v_\s^2 \right)^{-1/2}\ ,
\ee
and
\be
s =  - u_a a^a = \hat s \gamma_\s\ .
\ee

It is also natural to introduce the chemical potentials inferred by the observer;
\be
\mu_\x = -u^a \mu^\x_a = n_\x \B^\x + s \A^{\x\s}\ , 
\ee
and
\be
\Theta = s\B^\s + n_\p \A^{\p\s} + n_\e \A^{\e\s}\ , 
\ee

With these definitions it is straightforward to show that the total energy density measured by the observer will be
\be
\rho = u^a u^b T_{ab} = - \Psi + n_\p \mu_\p +n_\e \mu_\e + s\Theta \ , 
\ee
corresponding to the (integrated) first law of thermodynamics once we identify $\Psi$ as the generalized pressure and $\Theta$ as the temperature. 
Meanwhile, the momentum flux relative to the observer's frame is given by
\be
u^a T_{ab} = - \rho u_b -  n_\p \mu_\p v^\p_b - n_\e \mu_\e v^\e_b - s\Theta v^\s_b\ .
\ee
Here it is, first of all, natural to identify the heat flux as \cite{cesar1,cesar2}
\be
q^a = s \Theta v_\s^a\ .
\ee
We also see that we can choose the observer frame in such a way that this is the only relative momentum flux. To do this, we let
\be
\left(\rho+\Psi\right) v^a = n_\p \mu_\p v_\p^a + n_\e \mu_\e v_\e^a = 0 \ .
\ee
This is the natural extension of the ``centre of mass'' frame we used in the low-velocity model, cf., eq.~\eqref{vdef}. We also define the velocity difference (as before);
\be
w^a = v_\p^a - v_\e^a\ .
\ee
In the frame where $v^a=0$ (which will be assumed from now on) we have
\be
v_\p^a = {n_\e \mu_\e \over \rho + \Psi} w^a \ ,
\ee
and
\be
v_\e^a = -{n_\p \mu_\p \over \rho + \Psi} w^a \ ,
\ee
which means that the charge current can be written 
\be
j^a = e \left( n_\p - n_\e \right)u^a  + e {n_\p n_\e \over \rho+\Psi}\left( \mu_\p + \mu_\e\right) w^a\ .
\label{cc}\ee
At this point we recognize an important difference with respect to the low-velocity discussion. While we were naturally led to the assumption of charge neutrality in that case, the situation is much less clear now. This is immediately obvious from \eqref{cc} once we recall that the densities in the first term are measured by the chosen observer, not in the respective rest frames. Hence, it would not be appropriate to assume  that $n_\p = n_\e$ at this point.

\subsection{Friction and causal heat flow}

Having introduced the various ingredients, let us move on to the new aspect of the problem; the equation that governs the heat propagation. As we are treating the entropy as an additional fluid, it follows from the general analysis in \cite{livrev} that the thermal dynamics will be governed by its own momentum equation. We already know from \cite{cesar1,cesar2} that this will lead to an equation that contains the thermal relaxation that is required to ensure 
causality. However, as the entropy need not be conserved this momentum equation takes a slightly different form from those that govern the (individually conserved) material components. We have \cite{cesar1,cesar2}
\be
2 s^a  \nabla_{[a} \Theta_{b]}  + \Theta_b \Gamma_\s = f^\s_b \ ,
\label{smom}\ee
where 
\be
\nabla_a s^a = \Gamma_\s \ge 0\ ,
\ee
 in accordance with the second law.
 
Building on the analysis of the low-velocity case, we know that the overall conservation of energy and momentum requires
\be
\sum_\x f_\x^a = 0 \qquad \longrightarrow \qquad f_\s^a = - f_\p^a - f_\e^a\ .
\ee
The form of the ``force'' that acts on the entropy thus follows immediately from the forces on the charged components. Extending the low-velocity model, we will allow for resistivity due to scattering between both charged species ($\R$) and entropy ($\cS_\x$). Thus, we let the forces take the (still phenomenological) form
\be
f_\p^a = \perp_\p^{ab} \left( e\R n_b^\e + \cS_\p u^\s_b \right)\ , 
\ee
and
\be
f_\e^a = \perp_\e^{ab} \left( e\R n_b^\p + \cS_\e u^\s_b \right)\ , 
\ee
where we have used $s^a = \hat{s} u_\s^a$. 
Combining these expressions and expanding the projections, we arrive at
\begin{multline}
f_\s^a = - e\R\left[ n_\e - n_\p\gamma_\e^2 \left( 1 - v^b_\p v^\e_b \right)\right] \left( u^a + v_\e^a\right)
 - e\R\left[ n_\p - n_\e\gamma_\p^2 \left( 1 - v^b_\p v^\e_b \right)\right] \left( u^a + v_\p^a\right)\\
 - \cS_\p \left[ u_\s^a - \gamma_\s \gamma_\p u_\p^a \left( 1 - v^b_\p v^\s_b \right)\right] 
  - \cS_\e \left[ u_\s^a - \gamma_\s \gamma_\e u_\e^a \left( 1 - v^b_\e v^\s_b \right)\right] \ .
\end{multline}

Let us now return to \eqref{smom}, focussing on the entropy creation rate. Contracting the equation with the observer's four velocity we easily arrive at 
\be
\Theta \Gamma_\s = - u^b f^\s_b + 2 s u^b v_\s^a \nabla_{[a} \Theta_{b]} \ .
\label{Gs1}\ee
We need to constrain this in such a way that the right-hand side is non-negative. To do this, we first need the contraction between the entropy force and the four velocity. This leads to
\begin{multline}
- u^b f^\s_b = {e \R \over \rho + \Psi} \left( n_\e^2 \mu_\e \gamma^2_\p + n_\p^2 \mu_\p \gamma_\e^2 \right) w^2 + {\gamma_s \over \left( \rho + \Psi \right)^2 } \left[ \cS_\p \gamma_\p^2 \left( n_\e \mu_\e\right)^2 + \cS_\e \gamma_\e^2 \left( n_\p \mu_\p\right)^2 \right] w^2 \\
+ {\gamma_\s \over \rho+\Psi} \left( \cS_\e \gamma_\e^2 n_\p \mu_\p - \cS_\p \gamma_\p^2 n_\e \mu_\e\right) {w^b q_b \over s\Theta} \ .
\label{fric}\end{multline}
In this expression, the first two terms on the right-hand side will be positive as long as $\R\ge0$ and $\cS_\x\ge0$. The sign of the third term is not so clear. 

Moving on to the final term in \eqref{Gs1}, we first of all note that it is proportional to $v_\s^a\propto q^a$. Defining
\be
\beta_1 = {1 \over s \Theta }\left( \Theta - n_\p \A^{\p\s} - n_\e \A^{\e\s} \right) \ , 
\ee
and
\be
\beta_2 = {n_\p n_\e \over \rho+\Psi} \left( \mu_\e \A^{\p\s} - \mu_\p \A^{\e\s} \right)\ , 
\ee
we have
\be
\Theta_a = \Theta u_a + \beta_1 q_a + \beta_2 w_a\ , 
\ee
and we find that
\be
2 u^b \nabla_{[a} \Theta_{b]}  = - \perp_a^b \nabla_b \Theta - \Theta \dot{u}_a - \beta_1 \dot{q}_a - \beta_2 \dot{w}_a
- \dot\beta_1 q_a - \dot\beta_2 w_a - \left( \beta_1 q^b + \beta_2 w^b \right) \nabla_a u_b\ .
\label{gato}\ee

Combining \eqref{fric} and \eqref{gato} we see that $\Gamma_\s$ satisfies the required constraint provided that~\footnote{The strategy is to construct terms that are explicitly positive definite, e.g. quadratic in the relevant fluxes. The procedure is not necessarily unique, but the combinations that we have opted to use seem ``natural''.}
\begin{multline}
\kappa \beta_1 \dot q_a+ \left( 1 + \kappa \dot\beta_1 \right) q_a = - \kappa \Big[  \perp_a^b \nabla_b \Theta + \Theta \dot{u}_a + \beta_2 \dot{w}_a
+ \dot\beta_2 w_a + \left( \beta_1 q^b + \beta_2 w^b \right) \nabla_a u_b \\ 
-  {\gamma_\s \over s(\rho+\Psi)} \left( \cS_\e \gamma_\e^2 n_\p \mu_\p - \cS_\p \gamma_\p^2 n_\e \mu_\e\right) w_a \Big]\ , 
\end{multline}
with $\kappa \ge 0$. This has the form of a Cattaneo-type equation \cite{cesar1}, and a comparison of the $q_a$ and $\dot q_a$ term suggests that the  thermal relaxation time is
\be
\tau = {\kappa \beta_1 \over 1 + \kappa \dot\beta_1} \ .
\ee
However, the equation is also coupled to the four-acceleration $\dot u^a$ and the variation of the charge current, in terms of  $\dot w^a$, so if we want to infer the actual relaxation times in the problem we need to consider the coupled system. 

Combining the relevant contributions, we find that the total entropy creation rate is given by 
\begin{multline}
\Gamma_\s = {1\over \Theta } \Bigg[ {q^2 \over \kappa \Theta}  +  {e \R \over \rho + \Psi} \left( n_\e^2 \mu_\e \gamma^2_\p + n_\p^2 \mu_\p \gamma_\e^2 \right) w^2 \\
+ {\gamma_s \over \left( \rho + \Psi \right)^2 } \left[ \cS_\p \gamma_\p^2 \left( n_\e \mu_\e\right)^2 + \cS_\e \gamma_\e^2 \left( n_\p \mu_\p\right)^2 \right] w^2\Bigg] \ge 0 \ .
\end{multline}

\subsection{Ohm's law}

The derivation of the generalized form of Ohm's law follows the same steps as in the linear model, although now we need to keep careful track of the different redshift factors, etcetera. Basically, we want to construct the weighted difference between the momentum equations for the two charged components, but the two momenta now depend also on the entropy flux. In the frame associated with our chosen observer, we have 
\be
 \mu^\x_a = \mu_\x (u_a +v^\x_a ) + s \A^{\x\s} \left( v^\s_a - v^\x_a \right) \equiv \mu_\x (u_a +v^\x_a ) + \cW^\x_a \ ,
\ee
where we have introduced the convenient combinations
\be
\cW^\p_a = s\A^{\p\s} \left( {q_a \over s\Theta} - {n_\e\mu_\e\over\rho+\Psi} w_a \right)\ ,
\ee
and
\be
\cW^\e_a = s\A^{\e\s} \left( {q_a \over s\Theta} + {n_\p\mu_\p\over\rho+\Psi} w_a \right)\ ,
\ee
(obviously expressed in the chosen frame).

Given these expressions it follows that
\begin{multline}
f^\x_b = 2 n^a_\x \nabla_{[a} \mu^\x_{b]}  = n_\x \mu_\x \left( \dot u_b + \dot v^\x_b + v_\x^a \nabla_b u_a \right) \\
+ n_\x \perp^a_b \nabla_a \mu_\x + n_\x u^a v^\x_b \nabla_a \mu_\x +  2n_\x u^a \nabla_{[a} \cW^\x_{b]}  + 2n_\x v_\x^a \nabla_{[a} \mu^\x_{b]} \ .
\end{multline}

The weighted difference equation (inevitably rather complicated) combines three pieces. On the left-hand side we have
\begin{multline}
f^\mathrm{D}_b = {1\over n_\p \mu_\p} f^\p_b - {1\over n_\e \mu_\e} f^\p_e = \dot w_b + w^a\nabla_b u_a \\
+ \perp^a_b \left( {1\over \mu_\p} \nabla_a \mu_\p - 
 {1\over \mu_\e} \nabla_a \mu_\e \right) + u^a \left({1\over \mu_\p}v_\p^b  \nabla_a \mu_\p - 
 {1\over \mu_\e} v_\e^b \nabla_a \mu_\e \right)  \\
 + {2\over \mu_\p} u^a   \nabla_{[a} \cW^\p_{b]}  - {2 \over \mu_\e} u^a  \nabla_{[a} \cW^\e_{b]}  
 + {2 \over \mu_\p} v_\p^a  \nabla_{[a} \mu^\p_{b]}  - {2 \over \mu_\e} v_\e^a  \nabla_{[a} \mu^\e_{b]} \ .
\end{multline}

In the frame associated with the chosen observer, this expression takes the form~\footnote{For reasons of clarity we will not expand the terms that are quadratic in the relative velocities. It is straightforward to do so, but the final expressions are messy and not very instructive. The linear terms, on the other hand, highlight the explicit coupling between the different fluxes in the problem.}
\begin{multline}
f^\mathrm{D}_b = \dot{w}_b + w^a \nabla_a u_b + \perp^a_b\left( {1\over \mu_\p} \nabla_a \mu_\p - {1\over\mu_\e} \nabla_a \mu_\e \right) \\
+ {1 \over \mu_\p\mu_\e (\rho+\Psi)} 2 u_{(a}w_{b)} \left[ n_\e \mu_\e^2 \nabla^a \mu_\p - n_\p \mu_\p^2 \nabla^a \mu_\e \right] \\
-\perp^w_{ab} \left[ \left({n_\e \mu_\e \over \rho+\Psi} \right)^2 {1\over \mu_\p} \nabla^a \mu_\p - \left({n_\p \mu_\p \over \rho+\Psi} \right)^2 {1\over \mu_\e} \nabla^a \mu_\e \right] \\
+ u^a \left[ 2\nabla_{[a} \left( {\beta_4 \over s\Theta} q_{b]} \right) - 2\nabla_{[a}\left( \beta_3 w_{b]} \right)  - \cW^\p_b \nabla_a \left( {1\over\mu_\p} \right) + \cW^\e_b \nabla_a \left( {1\over\mu_\e} \right) \right] \\
+ 2\mathcal{D} w^a \nabla_{[a}w_{b]} 
- 2 {n_\e\mu_\e \over \rho+\Psi} w^a w_{[b}\nabla_{a]}\left(  {n_\e\mu_\e \over \rho+\Psi} \right) + 
2 {n_\p\mu_\p \over \rho+\Psi} w^a w_{[b}\nabla_{a]}\left(  {n_\p\mu_\p \over \rho+\Psi} \right) \\
+ 2 {n_\e\mu_\e \over\mu_\p (\rho+\Psi)} w^a \nabla_{[a}\cW^\p_{b]} -  2 {n_\p\mu_\p \over\mu_\e (\rho+\Psi)} w^a \nabla_{[a}\cW^\e_{b]}   \ .
\end{multline}
Here we have introduced
\begin{multline}
\beta_3 = {s \over \mu_\e  \mu_\p (\rho+\Psi)} \left( n_\e \mu_\e^2 \A^{\p\s} + n_\p \mu_\p^2 \A^{\e\s} \right) \\
= {s \over n_\p n_\e} \left[ {\Theta \over \rho+\Psi} \left( s \beta_1 -1 \right) + { n_\e \mu_\e -n_\p \mu_\p \over n_\p \mu_\p n_\e \mu_\e} \beta_2 \right] \ ,
\end{multline}
\be
\beta_4 = {s (\rho+\Psi) \over n_\e \mu_\e n_\p \mu_\p} \beta_2 \ .
\ee
and
\be
\mathcal{D} = \left({n_\e\mu_\e\over \rho+\Psi} \right)^2- \left({n_\p\mu_\p\over \rho+\Psi} \right)^2
\ee

We have also used the projection orthogonal to $w^a$;
\be
\perp^w_{ab} = w^2 g_{ab} - w_a w_b \ .
\ee
Meanwhile, the right-hand side is made up of, first of all, the combined friction forces;
\begin{multline}
f^\mathrm{R}_b = {1\over n_\p\mu_\p} \perp^\p_{ab} \left( e\R n_\e^a + \cS_\p u_\s^a \right) - 
{1\over n_\e\mu_\e} \perp^\e_{ab} \left( e\R n_\p^a + \cS_\e u_\s^a \right) \\
=
{e\R \over n_\p \mu_\p n_\e \mu_\e} \left[ {1\over \rho+\Psi} \left( n_\p^3 \mu_\p^2 \gamma_\e^2 - n_\e^3\mu_\e^2 \gamma_\p^2 \right) w^2 u_b 
- \left( n_\e^2 \mu_\e \gamma_\p^2 + n_\p^2 \mu_\p\gamma_\e^2 \right) \left( 1 - {s\Theta \over \rho+\Psi} \right) w_b\right] \\
+ \cS_\p {\gamma_\s \gamma_\p^2 \over n_\p\mu_\p} \Bigg\{ {n_\e\mu_\e\over \rho+\Psi} \left( {w^a q_a \over s\Theta} - {n_\e\mu_\e w^2 \over \rho+\Psi} \right) 
u_b + \left[ 1 - \left( {n_\e\mu_\e \over \rho+\Psi}\right)^2 w^2 \right]
{q_b \over s\Theta } \\
- {n_\e\mu_\e \over \rho+\Psi} \left( 1 - {n_\e \mu_\e \over \rho+\Psi} {w^a q_a \over s\Theta} \right) w_b \Bigg\} \\
+ \cS_\e {\gamma_\s \gamma_\e^2 \over n_\e\mu_\e} \Bigg\{ {n_\p\mu_\p\over \rho+\Psi} \left( {w^a q_a \over s\Theta} + {n_\p\mu_\p w^2 \over \rho+\Psi} \right) 
u_b - \left[ 1 - \left( {n_\p\mu_\p \over \rho+\Psi}\right)^2 w^2 \right]
{q_b \over s\Theta } \\
- {n_\p\mu_\p \over \rho+\Psi} \left( 1 + {n_\p \mu_\p \over \rho+\Psi} {w^a q_a \over s\Theta} \right) w_b \Bigg\} \ , 
\end{multline}
where the second equality holds in the frame where $v^a=0$. 

The final part accounts for the electromagnetic field. We need
\begin{multline}
f^\mathrm{M}_b = {1\over n_\p\mu_\p} en_\p^a F_{ba} + {1\over n_\e\mu_\e} en_\e^a F_{ba} = e {\mu_\p+\mu_\e \over \mu_\p\mu_\e} u^a F_{ba} 
+ e\left( {1 \over \mu_\p} v_\p^a + {1\over \mu_\e} v_\e^a\right) F_{ba} \\
= e \left( {\mu_\p+\mu_\e \over \mu_\p\mu_\e}\right) E_b + e{n_\e \mu_\e^2 - n_\p \mu_\p^2 \over \mu_\p \mu_\e \left( \rho+\Psi\right)}\left[ u_b \left(w^a E_a\right) +\epsilon_{bac} w^a B^c \right] \ .
\end{multline}
Again the second equality holds only in the chosen frame.

The final relation follows from the combination \eqref{fbal} after projecting out the component orthogonal to $u^a$. The result is (inevitably) rather complex and may not be particularly instructive. Yet, we provide it in the interest of completeness. The generalised form for Ohm's law for a hot two-component plasma can be written (expressed in terms of $w^a$ rather than the spatial component of the charge current $J^a$, for convenience) 
\begin{multline}
 \dot{w}_b + w^a \nabla_a u_b + \perp^a_b\left( {1\over \mu_\p} \nabla_a \mu_\p - {1\over\mu_\e} \nabla_a \mu_\e \right) \\
+ {1 \over \mu_\p\mu_\e (\rho+\Psi)} w_b u^a  \left[ n_\e \mu_\e^2 \nabla_a \mu_\p + n_\p \mu_\p^2 \nabla_a \mu_\e \right] \\
-\left( w^2 \perp^a_b - w^a w_b \right) \left[ \left({n_\e \mu_\e \over \rho+\Psi} \right)^2 {1\over \mu_\p} \nabla_a \mu_\p - \left({n_\p \mu_\p \over \rho+\Psi} \right)^2 {1\over \mu_\e} \nabla_a \mu_\e \right] \\
+ u^a \left[ 2\nabla_{[a} Q_{b]}  - \cW^\p_b \nabla_a \left( {1\over\mu_\p} \right) + \cW^\e_b \nabla_a \left( {1\over\mu_\e} \right) \right] \\
+ \perp_b^c \left[  \mathcal{D} w^a \nabla_a w_c  - \nabla_c \left( \mathcal{D} w^2 \right)\right]  - {1 \over 2} w^a w_b \nabla_a \mathcal{D} \\
+ {2 \over \rho+\Psi} \perp_b^c \left(  {n_\e\mu_\e \over\mu_\p } w^a \nabla_{[a}\cW^\p_{c]} -  {n_\p\mu_\p \over\mu_\e} w^a \nabla_{[a}\cW^\e_{c]} 
\right) \\
= - {e\R \over n_\p \mu_\p n_\e \mu_\e}  \left( n_\e^2 \mu_\e \gamma_\p^2 + n_\p^2 \mu_\p\gamma_\e^2 \right) \left( 1 - {s\Theta \over \rho+\Psi} \right) w_b\\
+ \cS_\p {\gamma_\s \gamma_\p^2 \over n_\p\mu_\p} \Bigg\{  \left[ 1 - \left( {n_\e\mu_\e \over \rho+\Psi}\right)^2 w^2 \right]
{q_b \over s\Theta } 
- {n_\e\mu_\e \over \rho+\Psi} \left( 1 - {n_\e \mu_\e \over \rho+\Psi} {w^a q_a \over s\Theta} \right) w_b \Bigg\} \\
- \cS_\e {\gamma_\s \gamma_\e^2 \over n_\e\mu_\e} \Bigg\{  \left[ 1 - \left( {n_\p\mu_\p \over \rho+\Psi}\right)^2 w^2 \right]
{q_b \over s\Theta } 
+ {n_\p\mu_\p \over \rho+\Psi} \left( 1 + {n_\p \mu_\p \over \rho+\Psi} {w^a q_a \over s\Theta} \right) w_b \Bigg\} \\
+ e \left( {\mu_\p+\mu_\e \over \mu_\p\mu_\e}\right) E_b +  e{n_\e \mu_\e^2 - n_\p \mu_\p^2 \over \mu_\p \mu_\e \left( \rho+\Psi\right)}\epsilon_{bac} w^a B^c \ ,
\end{multline}
where we have defined 
\be
Q_b = {\beta_4 \over s\Theta} q_{b} - \beta_3 w_{b} \ .
\ee

\subsection{The total momentum equation}

As in the two-component system, the model is completed by the total momentum equation, which follows (more or less) immediately from the divergence of the stress-energy tensor. In the frame moving with $u^a$ (where $v^a=0$), the matter stress-energy tensor takes the form
\be
T^\mathrm{M}_{ab} = \rho u_a u_b + \perp_{ab} \Psi + 2 u_{(a} q_{b)} + \alpha w_a w_b + {2 \beta_2 \over \Theta} w_{(a}q_{b)} + {\beta_1 \over \Theta} q_a q_b \ ,
\ee
with
\begin{multline}
\alpha = {n_\e \mu_\e n_\p \mu_\p \over \rho + \Psi} \left( 1 - {s\Theta \over \rho+\Psi} \right) - \left( {n_\e \mu_\e \over \rho+\Psi }\right)^2 s n_\p \A^{\p\s} 
-  \left( {n_\p \mu_\p \over \rho+\Psi }\right)^2 s n_\e \A^{\e\s} \\
= {n_\e \mu_\e n_\p \mu_\p \over \rho + \Psi} \left( 1 - {s^2 \beta_1 \Theta \over \rho+\Psi}   \right) - {s \left( n_\e \mu_\e - n_\p\mu_\p\right) \over \rho + \Psi}  \beta_2 \ .
\end{multline}

As usual, $\nabla_a T^{ab}=0$ can be divided into a component along the four velocity and an orthogonal piece. After a bit of algebra, recalling that the electromagnetic contribution is given by eq.~\eqref{loren}, we find that 
the former can be written
\begin{multline}
\dot\rho + (\rho+\Psi) \nabla_a u^a + \nabla_a q^a \\ - u_b \left[ \dot q^b + \alpha w^a \nabla_a w^b
 + {\beta_2 \over \Theta} \left( w^a \nabla_a q^b + q^a \nabla_a w^b \right)  + {\beta_1 \over \Theta} q^a \nabla_a q^b \right] \\
  = 
 e {n_\p n_\e \over\rho+\Psi} \left( \mu_\p + \mu_\e\right) \left(w_a E^a\right)\ .
\end{multline}
Meanwhile, the orthogonal projection leads to the momentum equation 
\begin{multline}
\left( \rho + \Psi \right)\dot u^b  + \perp^{ab} \nabla_a \Psi + q^a \nabla_a u^b  +q^b \nabla_a \left( u^a + V_1^a \right) \\
+ \perp^b_c \left( \dot q^c + V_2^a \nabla_a w^c + V_1^a \nabla_a q^c \right) + w^b \nabla_a V_2^a \\
=  e \left( n_\p - n_\e \right) E^b + e {n_\p n_\e \over\rho+\Psi} \left( \mu_\p + \mu_\e\right) \epsilon^{bac} w_a B_c\ ,
\end{multline}
where we have defined
\be
V_1^a = {1\over \Theta} \left(  \beta_2 w^a + \beta_1 q^a \right)\ ,
\ee
and
\be
V_2^a = \alpha w^a + {\beta_2 \over\Theta} q^a \ .
\ee

\subsection{A linearised model}

The final equations for the coupled three-component model are obviously rather complex. This is not surprising since, apart from working in a specific observer frame, we did not make any simplifications. Hence, the model is quite general, including the relevant nonlinearities and redshift factors. The main take home message should be that the steps involved in the derivation are natural and intuitive, but the expressions involved will be messy. One may  query the immediate usefulness of the analysis, as it takes us far beyond what is currently considered in applications. However, the argument on behalf of the defence is clear. Once we have worked our way through the general case it is relatively straightforward to reduce the complexity by considering specific models. This is, in fact,  a valuable exercise as it provides a clearer insight into the key features of the hot system. 

A natural, and in many cases of interest reasonable, assumption is that we only need to retain the linear relative velocities.  As in the cold model, we neglect higher order terms in $w^a$ and $q^a$, and we also ignore all the redshift factors by taking $\gamma_\x\approx 1$. It then follows naturally that the pressure is $\Psi=P$ and  the temperature is $\Theta = T$.  In the interest of clarity we will also ignore the resistive scattering between entropy (phonons) and the material components, i.e. we set $\cS_\x=0$. These assumptions lead to, 
i) the heat equation;
\begin{multline}
\kappa \beta_1 \dot q_a+ \left( 1 + \kappa \dot\beta_1 \right) q_a = - \kappa \left[  \perp_a^b \nabla_b T + T \dot{u}_a + \beta_2 \dot{w}_a
+ \dot\beta_2 w_a + \left( \beta_1 q^b + \beta_2 w^b \right) \nabla_a u_b \right]\ ,
\label{heateq}\end{multline}
ii) the generalised version of Ohm's law;
\begin{multline}
e \left( {\mu_\p+\mu_\e \over \mu_\p\mu_\e}\right) E_b -  e{ n_\p \mu_\p^2- n_\e \mu_\e^2  \over \mu_\p \mu_\e \left( P+\rho\right)}\epsilon_{bac} w^a B^c 
- {e\R \over n_\p \mu_\p n_\e \mu_\e}  \left( n_\e^2 \mu_\e+ n_\p^2 \mu_\p \right) \left( 1 - {sT \over P+\rho} \right) w_b\\=
 \dot{w}_b + w^a \nabla_a u_b + \perp^a_b\left( {1\over \mu_\p} \nabla_a \mu_\p - {1\over\mu_\e} \nabla_a \mu_\e \right) \\
+ {1 \over \mu_\p\mu_\e (P+\rho)} w_b u^a \left[ n_\e \mu_\e^2 \nabla_a \mu_\p + n_\p \mu_\p^2 \nabla_a \mu_\e \right] \\
+ u^a \left[ 2\nabla_{[a} Q_{b]}  - \cW^\p_b \nabla_a \left( {1\over\mu_\p} \right) + \cW^\e_b \nabla_a \left( {1\over\mu_\e} \right) \right] \ ,
\label{ohmeq}\end{multline}
and, 
iii) the total momentum conservation equation;
\begin{multline}
\left( P+\rho \right)\dot u^b  + \perp^{ab} \nabla_a P + q^a \nabla_a u^b  +q^b \nabla_a u^a  
+ \perp^b_c \dot q^c\\
=  e \left( n_\p - n_\e \right) E^b + e {n_\p n_\e \over P+\rho} \left( \mu_\p + \mu_\e\right) \epsilon^{bac} w_a B_c\ .
\label{momeq}\end{multline}

We simplify these relations further by noting that we can reinstate the assumption of charge neutrality, as the issues alluded to after eq.~\eqref{cc} originate from the redshift factors. Thus we let $n_\p=n_\e$, which leads to a number of simplifications (the arguments are the same as in the cold case). Focussing on the proton-electron plasma, we also assume that $\mu_\e \ll \mu_\p$.  It follows that the charge current is (again) given by
\be
J^a = e n_\e w^a \ .
\ee
We can use this to write \eqref{heateq} in the elegant form  
\be
q_a = - \kappa \left(  \perp_a^b \nabla_b T + T \dot{u}_a + 2 u^b \nabla_{[b} \tilde{Q}_{a]} \right)\ ,
\ee
where
\be
 \tilde{Q}_a = \beta_1 q_a + {\beta_2 \over e n_\e } J_a \ .
\ee
The thermal relaxation is encoded in this quantity.

Turning to the momentum equation \eqref{momeq}, we have
\be
\left( P+\rho \right)\dot u^b  + \perp^{ab} \nabla_a P + q^a \nabla_a u^b  +q^b \nabla_a u^a  
+ \perp^b_c \dot q^c\\
=   \epsilon^{bac} J_a B_c\ .
\ee

Finally, we find that Ohm's law simplifies to~\footnote{Here we have neglected the thermal pressure, i.e. we have assumed that 
$s\Theta \ll P+\rho$. It is natural to make this assumption along with $\mu_e \ll \mu_\p$, and moreover the thermal pressure  is likely to be negligible in many realistic situations.}
\begin{multline}
E_b -{1\over e n_\e}  \epsilon_{bcd}J^c B^d - {\R \over e n_\e} J_b \\
= {\mu_\e \over e^2 n_\e} \left[ \perp_{ab} \dot{J}^a + J^a  \left(  \sigma_{ab} + \omega_{ab} + {4 \over 3} \theta \perp_{ab} \right)   \right] - {1 \over e} \perp^a_{\ b} \nabla_a \mu_\e + 2u^a \nabla_{[a}{Q}_{b]}\ .
\end{multline}

At this point, we have stripped the hot plasma model down to the level where it is easy to compare the final expressions to those of the cold model. At the linear level, the only difference is in the presence of the couplings that arise due to the entropy entrainment (expressed in terms of the different $\beta$ coefficients) and the explicit presence of the heat flux $q^a$ in the momentum equation. These may seem like minor adjustments, but they are significant. In particular, we need to retain the relevant relaxation times in order to ensure that the model is causal. 

Of course, the main differences between the two models we have developed enters at the nonlinear level (in the relative velocities).  At quadratic order, the problem is non-adiabatic (as $\Gamma_\s \neq 0$) and it is no longer natural to assume charge neutrality. Given these effects, it would be very interesting to study a quadratic model in more detail. However, the corresponding problem is somewhat involved so we prefer to postpone discussion of it for the future. 

The model is completed by three scalar relations (whose origin are the conservation laws for the fluxes). In the linear case, these take the simple form
\be
\dot\rho + (P+\rho) \theta + \nabla_a q^a - u_b \dot q^b = 0  \ , 
\ee
\be
\dot s + s \theta + \nabla_a \left( {q^a \over T} \right) = 0 \ , 
\ee
and 
\be
\nabla_a J^a = 0 \ , 
\ee
which can be replaced by 
\be
\dot n_\e + n_\e \theta = 0 \ .
\ee

\section{Concluding remarks}

We have developed the theory for charged fluids coupled to an electromagnetic field in the framework of general relativity, accounting for both a phenomenological resistivity and the relaxation times (associated with the charge current and the heat flux) that are required to ensure causality. The final formalism can be applied to a range of interesting problems in astrophysics and cosmology. The cold two-component plasma model (from Section~III) extends the ``ideal'' magnetohydrodynamics framework in several  directions, and the hot model (from Section~IV) adds dimensions that come into play when thermal aspects of the problem cannot be neglected. These developments are important as a number of interesting problems may require ``non-ideal'' aspects for their solution. Of most obvious relevance  are problems involving not only electromagnetic fields but the live spacetime of general relativity. Several key gravitational-wave sources come to mind, like core-collapse supernovae \cite{kota1} and compact binary mergers 
\cite{merge1,merge2}. Both cases involve strong gravity, a significant thermal component and magnetic fields. To apply a resistive framework to these problems is, of course, seriously challenging but this does not mean that we should not have aspirations in this direction \cite{wata,pale,taka}. Actual multi-fluid simulations \cite{zani} are also of obvious relevance. 

Focussing on relativistic stars, one can think of a number of unresolved problems, ranging from the dynamics of the magnetosphere and the pulsar emission mechanism to the formation and evolution of the star's interior magnetic field. These are problems where there has been significant progress, but further effort is required. In the case of the magnetosphere, the main focus has been on force-free models, but recent arguments \cite{spit} point to the need to include resistivity in the discussion. In the case of the formation and evolution of a compact star's 
global magnetic field, we need a better understanding of dynamo effects that may come into operation (see \cite{dynamo} and also \cite{brandenburg} for a recent review) and we also need
to understand the coupled evolution of the star's spin, temperature and magnetic field \cite{pons}. There are some very difficult issue to resolve here.

In fact, the suggested examples highlight the need to develop the theory further. Typical questions that would need to be addressed involve (i) the dynamics predicted by the model, e.g.  causality and stability of wave propagation and relation to issues like pulsar emission or the launch of outflows and jets, (ii) transitions between spatial regions where different simplifying assumptions are valid, such as a region in the magnetosphere where the fluid model applies and a low density region where the description breaks down and one would need to fall back on a kinetic theory model \cite{mark,meier,gedalin}, the transition from magnetosphere to interior field at the star's surface or, indeed, accreting systems where an ion-electron plasma describes the inflowing matter while regions in the magnetosphere may still be appropriately modelled as a pair-plasma, (iii) the role of more complex physics, like the superconductor that is expected to be present in the star's core \cite{supercon} or regions where the assumption that the medium is electromagnetically ``passive'' does not apply, possibly in the pasta region near the crust-core transition. The present work provides a foundation for developments in all these directions, but each problem is associated with specific challenges  that will need to be addressed if we want to make further  progress.

\acknowledgements

I would like to thank John Miller, Greg Comer, Bernard Schutz and Kostas Glampedakis for useful discussions. I am also pleased to acknowledge financial support from STFC in the UK (through grant number ST/J00135X/1).

\end{document}